\documentclass[review]{elsarticle}

\usepackage{amsmath}
\usepackage{amssymb}
\usepackage{hyperref}
\usepackage{graphics}      % standard graphics specifications
\usepackage{graphicx} 
\usepackage{bm}
\usepackage[english]{babel}
\usepackage[utf8]{inputenc}
\journal{Journal of \LaTeX\ Templates}

\begin{document}

\begin{frontmatter}

\title{Thermal entanglement in the mixed-spin Ising-Heisenberg double sawtooth frustrated ladder}

\author{H. Arian Zad\corref{mycorrespondingauthor}}
\address{Young Researchers and Elite Club, Mashhad Branch, Islamic Azad University, Mashhad, Iran}
\ead{arianzad.hamid@mshdiau.ac.ir}

\author{N. Ananikian}
\ead{ananik@mail.yerphi.am}

\address{Alikhanyan National Science Laboratory, Alikhanian Br. 2, 0036 Yerevan, Armenia}

\begin{abstract}
The entanglement between spin-1/2 interstitial Heisenberg dimers in the mixed spin-(1,1/2) Ising-XXZ  double sawtooth ladder is investigated at low temperature. Here, we consider a cyclic four-spin exchange interaction in square plaquette of each block, and investigate the effects of this amazing interaction on the bipartite entanglement between spin-1/2 interstitial dimers. Interestingly, we observe a remarkable difference in concurrence behavior with respect to the cyclic four-spin exchange interaction and magnetic field. Also, the critical points at which the concurrence vanishes are changed versus alteration of the anisotropic parameter of the interstitial Heisenberg dimers.

\end{abstract}

\begin{keyword}
Entanglement, double sawtooth,  cyclic four-spin exchange interaction
\MSC[2010] 00-01\sep  99-00
\end{keyword}

\end{frontmatter}

%\linenumbers

\section{Introduction}
There are quite lots of reasons to investigate the entanglement \cite{Horodecki2009,Wootters1998,Zhang2007,Langari2008,Saif2010,Vedral2001,Kamta,Ananikian2012,Arian2} and bipartite correlations \citep{Werlang2010,Abliz2011326,Sarandy2012} for the Heisenberg models, where they have been the main subject of numerous papers for a long period of time. for instance, There are many applications of the entanglement in quantum teleportation \cite{Zhang2007}, quantum NMR and quantum information processing \cite{Osterloh1,Furman2012,Dolde,ArianNMR}.

Besides simple 1-D Heisenberg spin chains, a wide variety of the spin models such as diamond chains \cite{Rojas2,Gu,Ananikian2012,Abgaryan1,Abgaryan2,Strecka1}, spin ladders \cite{Ivanov1,Strecka} and sawtooth ladders have been encountered with a great attention from theoretical and experimental view points \cite{Buttner,Bacq,Blundell,Arian1}.

\begin{figure}
\begin{center}
\includegraphics[width=6cm,height=5.5cm]{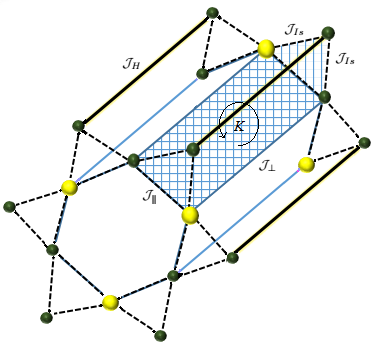}
\caption{Schematic structure of a mixed-spin Ising-XXZ Heisenberg  double sawtooth ladder with geometric frustration and additional ring exchange for $N=24$. }
\label{fig:figure1}
\end{center}
\end{figure}

In solid state physics and condensed matter science, the main goal of numerous papers is providing the exact solution for the various generalized versions of the mixed-spin (1,1/2) Ising-Heisenberg chains, which bring a deep insight into how the thermal and the magnetic properties are dependent on the spins-1/2 and spin-1 ordering \citep{Arian2,Abgaryan1,Ivanov1,Lisnyi,Verkholyak}. Recently, we investigated fermionic Ising-XXZ Heisenberg double sawtooth ladder in Ref. \cite{Arian1}, also  mixed-spin (1,1/2) Ising-XXZ double sawtooth ladder in Ref.  \cite{ArianC2}. Phase transitions and some thermodynamic parameters such as specific  heat, the magnetization and magnetic susceptibility have been numerically investigated in detail.
In the present work, we are going to study the thermal entanglement between the interstitial dimer Heisenberg half-spins of the mixed-spin (1,1/2) Ising-XXZ double sawtooth ladder where the nodal sites have Ising spins including both spin-1/2 and spin-1 on the legs in the presence of an external magnetic field at low temperature. The mixed-spin ladder of 24-spins (as an example) with periodic boundary conditions is shown in Fig. \ref{fig:figure1} . The number of spins in the ladder are selected even and the suggested procedure in this paper is used for $N\geq 8$.
For a chain with $N>8$, the number of spins will grow as $N+4$, i.e., $N\in\{8,12,16,20,24\cdots\}$.

The paper is organized as the following: in the next section we introduce the suggested model. In section \ref{TM}, we present the thermodynamic solution of the model using the transfer-matrix formalism approximately. In section \ref{Concurrence}, we have numerically discussed the correlation functions and the thermal concurrence. Section \ref{conclusions} is devoted to draw the conclusions.
\section{Model and method}\label{Model}
As we introduced the mixed-spin Ising-XXZ Heisenberg double sawtooth ladder with an analytical Hamiltonian in Ref. \cite{ArianC2}, here we explain the model briefly. The $i$-th block Hamiltonian  ${h}_{i}$ of the antiferromagnetic mixed spin double sawtooth ladder under the periodic boundary conditions (shaded region in Fig. \ref{fig:figure1}) can be given by
\begin{equation}\label{HamiltonianT}
\begin{array}{lcl}
{h}_{i}=\big[\mathcal{J}_{x}\big({\sigma}^{x}_{i,2}{\sigma}^{x}_{i,5}+{\sigma}^{y}_{i,2}{\sigma}^{y}_{i,5}\big)+\Delta{\sigma}^{z}_{i,2}{\sigma}^{z}_{i,5}\big]\\
+\mathcal{J}_{\parallel}\big({J}^{z}_{i,1}{\sigma}^{z}_{i,3}+{J}^{z}_{i,4}{\sigma}^{z}_{i,6}\big)+\frac{\mathcal{J}_{\perp}}{2}\big({J}^{z}_{i,1}{\sigma}^{z}_{i,6}+{\sigma}^{z}_{i,3}{J}^{z}_{i,4}\big)\\
+\mathcal{J}_{Is}\big[{J}^{z}_{i,1}{\sigma}^{z}_{i,2}+{\sigma}^{z}_{i,2}{\sigma}^{z}_{i,3}+{J}^{z}_{i,4}{\sigma}^{z}_{i,5}+
{\sigma}_{i,5}{\sigma}^{z}_{i,6}\big]+K P_{1346}^{i\circlearrowleft}\\
-\frac{{B}^{\prime}_{z}}{2}\big({J}^{z}_{i,1}+{\sigma}^{z}_{i,3}+{J}^{z}_{i,4}+{\sigma}^{z}_{i,6}\big)-{B}^{\prime\prime}_{z}\big({\sigma}^{z}_{i,2}+{\sigma}^{z}_{i,5}\big),
\end{array}
\end{equation}
where  ${\sigma^{\alpha}}=\lbrace {\sigma}^x, {\sigma}^y, {\sigma}^z \rbrace$ are Pauli operators (with $\hbar=1$), and ${ J}^{z}=diag(1, 0, -1)$ is spin-1 operator. ${B}^{\prime}_z$ and ${B}^{\prime\prime}_z$ are applied homogeneous magnetic fields  in the $z$-direction.  
$\mathcal{J}_{\perp}$ and $\mathcal{J}_{\parallel}$ are the bilinear exchange couplings on the rungs and along the legs of the plaquette in unit block, respectively. $K$ is the cyclic four-spin exchange interaction per plaquette, and $\mathcal{J}_{Is}$ is the Ising interaction between the spins on the legs of the block$^,$s plaquette and two interstitial Heisenberg dimer spins. 

The cyclic four-spin permutation operator for which all interactions are kind of Ising exchange interaction can be written as the product of three transposition operators $P_{1346}^{\circlearrowleft}=P_{13}^{\circlearrowleft}P_{14}^{\circlearrowleft}P_{16}^{\circlearrowleft}$ where $P_{13}^{\circlearrowleft}=1/2(1+{J}^{z}_{1}\sigma^{z}_{3})$, hence 
\begin{equation}\label{FourC}
\begin{array}{lcl}
P_{1346}^{i\circlearrowleft}=\frac{1}{8}\big[1+{J}^{z}_{1}\sigma^{z}_{3}+{J}^{z}_{1}{J}^{z}_{4}+{J}^{z}_{1}\sigma^{z}_{6}+
({J}^{z}_{1}\sigma^{z}_{3})\cdot({J}^{z}_{1}{J}^{z}_{4})+\\
({J}^{z}_{1}\sigma^{z}_{3})\cdot({J}^{z}_{1}\sigma^{z}_{6})+({J}^{z}_{1}{J}^{z}_{4})\cdot({J}^{z}_{1}\sigma^{z}_{6})+
({J}^{z}_{1}\sigma^{z}_{3})\cdot({J}^{z}_{1}{J}^{z}_{4})\cdot({J}^{z}_{1}\sigma^{z}_{6})\big],
\end{array}
\end{equation}
 which contains both bilinear and biquadratic terms of the spin-1/2 and spin-1 operators. Note that here, all of introduced parameters are considered dimensionless.
\section{Approximate solution in the transfer matrix formalism}\label{TM}
 The interstitial Heisenberg dimer coupling can be expressed as
\begin{equation}\label{density matrices}
\big({\boldsymbol\sigma}_{i,2}\cdot{\boldsymbol\sigma}_{i,5}\big)_{\Delta,\mathcal{J}_{x}}= \left(
\begin{array}{cccc}
\frac{\Delta}{4} & 0 & 0 & 0 \\
0 & -\frac{\Delta}{4} & \frac{\mathcal{J}_x}{2} & 0 \\
0 &\frac{\mathcal{J}_x}{2} & -\frac{\Delta}{4}  & 0\\
 0 & 0 & 0 & \frac{\Delta}{4}
\end{array} \right).
\end{equation}
The partition function of the ladder is given by
%\begin{equation}\label{PF}
$\mathcal{Z}=Tr\Big[\displaystyle\prod_{i=1}^{M}\exp(-\beta h_{i})\Big],$
%\end{equation}
where $\beta=\frac{1}{k_{B}T}$, $k_{B}$ is the Boltzmann’s constant and $T$ is the temperature.
In terms of the qubit-qutrit standard basis of the composite spin operators $\{ J_{i,1}^{z},\sigma_{i,6}^{z},\sigma_{i,3}^{z},J_{i,4}^{z} \} $ on the two consecutive rungs of the  plaquette in block $i$, the partition function $\mathcal{Z}$ can be defined as
\begin{equation}\label{PF}
\begin{array}{lcl}
\mathcal{Z}=Tr\big[ \langle J_{1,1}^z\sigma_{1,6}^z\vert\mathcal{T}\vert \sigma_{1,3}^z J_{1,4}^z\rangle\langle \sigma_{2,1}^z
J_{2,6}^z\vert \mathcal{T}\vert J_{2,3}^z\sigma_{2,4}^z\rangle
\cdots\langle \sigma_{M,1}^z J_{M,6}^z\vert \mathcal{T}\vert J_{M,3}^z\sigma_{M,4}^z\rangle \big],
\end{array}
\end{equation}
where $\sigma_{i,j}^z=\pm1$ and $J_{i,j}^z=\pm1, 0$ and
\begin{equation}\label{TrM}
\begin{array}{lcl}
\mathcal{T}(i)=\langle J_{i,1}^z\sigma_{i,6}^z\vert\exp(-\beta h_{i})\vert \sigma_{i,3}^z J_{i,4}^z\rangle=
\sum\limits_{k=1}^4\exp\big[-\beta\mathcal{E}_k(J_{i,1}^z\sigma_{i,6}^z,\sigma_{i,3}^z J_{i,4}^z)\big].
\end{array}
\end{equation}
Four eigenvalues of the $i-$th block with Hamiltonian $h_{i}$ are
\begin{equation}\label{eigenvalues1}
\begin{array}{lcl}
\mathcal{E}_1(i)={\Delta}+\mathcal{J}_{Is}\big(J_{i,1}^z+\sigma_{i,3}^z+J_{i,4}^z+\sigma_{i,6}^z\big)+\Xi-2B,\\
\mathcal{E}_2(i)=\\ 
-{\Delta}+\Xi+\sqrt{\mathcal{J}_{Is}^2\big(J_{i,1}^z+\sigma_{i,3}^z-J_{i,4}^z-\sigma_{i,6}^z\big)^2+4\mathcal{J}_{x}^2},\\
\mathcal{E}_3(i)=\\ 
-{\Delta}+\Xi-\sqrt{\mathcal{J}_{Is}^2\big(J_{i,1}^z+\sigma_{i,3}^z-J_{i,4}^z-\sigma_{i,6}^z\big)^2+4\mathcal{J}_{x}^2},\\
\mathcal{E}_4(i)={\Delta}-\mathcal{J}_{Is}\big(J_{i,1}^z+\sigma_{i,3}^z+J_{i,4}^z+\sigma_{i,6}^z\big)+\Xi+2B,
\end{array}
\end{equation}
where 
\begin{equation}\label{Xi}
\begin{array}{lcl}
\Xi=\mathcal{J}_{\parallel}\big(J_{i,1}^{z}\sigma_{i,3}^{z}+J_{i,4}^{z}\sigma_{i,6}^{z}\big)+\frac{\mathcal{J}_{\perp}}{2}\big(J_{i,1}^{z}\sigma_{i,6}^{z}+\sigma_{i,3}^{z}J_{i,4}^{z}\big)+\\
\frac{K}{8}\big[J_{i,1}^{z3}\sigma_{i,3}^{z}J_{i,4}^{z}\sigma_{i,6}^{z}+J_{i,1}^{z2}\sigma_{i,3}^{z}J_{i,4}^{z}+
J_{i,1}^{z2}J_{i,4}^{z}\sigma_{i,6}^{z}+\\
J_{i,1}^{z2}\sigma_{i,3}^{z}\sigma_{i,6}^{z}+J_{i,1}^{z}J_{i,4}^{z}+J_{i,1}^{z}\sigma_{i,3}^{z}+J_{i,1}^{z}\sigma_{i,6}^{z}+1\\
-B\big(J_{i,1}^{z}+\sigma_{i,3}^{z}+J_{i,4}^{z}+\sigma_{i,6}^{z}\big)-2B\big].
\end{array}
\end{equation}
The components of the transfer matrix can be obtained by using eigenvalues (\ref{eigenvalues1}).

%Since the mixed spin double sawtooth ladder is translational invariant and all of $h_i$ are  independent of the site $i$,
Due to the commutation relation between different block Hamiltonians, $[{h}_{i},{h}_{j}]=0$, equation (\ref{PF}) can be expressed as
%\begin{equation}\label{ConvertedZ}
$\mathcal{Z}=Tr\big[\mathcal{T}^M\big].$
%\end{equation}
The partition function of the model can be expressed through six eigenvalues of the transfer matrix $\mathcal{T}(i)$ as
%\begin{equation}\label{TotalZ}
$\mathcal{Z}=\Lambda_1^{M}+\Lambda_2^{M}+\Lambda_3^{M}+\Lambda_4^{M}+\Lambda_5^{M}+\Lambda_6^{M}.$
%\end{equation}
 In the thermodynamic limit, only the largest eigenvalue $\Lambda_{max}$ is acquired to gain the partition function. 

Since, all elements of the transfer matrix are kinds of non-zero long term polynomials, obtaining the eigenvalues of the real transfer matrix 
$\mathcal{T}(i)$ is a big challenge for the available processors, hence, its quite interest to follow the calculations in the framework of an approximate procedure to extract the largest eigenvalue of the transfer matrix. We here explain this procedure with extensive uses in detail. By  investigating the mixed-spin double sawtooth ladder numerically, such that we examine the transfer matrix elements for the ranges 
 $\mathcal{J}_{\parallel}=\mathcal{J}_{Is}=1$, 
$\mathcal{J}_{\perp}=rand(1..8)$, $\mathcal{J}_{x}=rand(-2..2)$, $\Delta=rand(-1..6)$, $\beta=rand(0..3)$ and
 $B=rand(0..4)$ over than 100-times randomly calculating,  we found that for the determined ranges of the Hamiltonian parameters, several components of the transfer matrix are almost effectless and they can be neglected. So, we construct a new transfer matrix 
$\mathcal{T}^{\prime}(i)$ with less non-zero components including the most effective components of the real transfer matrix $\mathcal{T}(i)$. The new transfer matrix can be given by
\begin{equation}\label{NewTmat}
\mathcal{T}^{\prime}(i) = \left(
\begin{array}{cccccc}
 \mathcal{T}_{11} &  \mathcal{T}_{12} &  \mathcal{T}_{13} &  0 &  \mathcal{T}_{15} & 0 \\
\mathcal{T}_{21} &  \mathcal{T}_{22} &  0 &  0 &  0 & 0 \\
\mathcal{T}_{31} &  0 &  \mathcal{T}_{33} & 0 &  \mathcal{T}_{35} & 0 \\
0 &  0 &  0 & 0 &  0 & 0 \\
\mathcal{T}_{51} &  0 &  \mathcal{T}_{53} &  0 &  \mathcal{T}_{55} & 0 \\
0 &  0 &  0 &  0 &  0 & 0 \\
\end{array} \right).
\end{equation}
 Interestingly, the largest eigenvalue of the new transfer matrix $\mathcal{T}^{\prime}(i)$ is equivalent to the largest eigenvalue of the real transfer matrix $\mathcal{T}(i)$ with high accuracy.
 Free energy Gibbs per block for infinite chain with respect to the largest eigenvalue can be written as \cite{Paulinelli}
 \begin{equation}\label{FreeE}
 \begin{array}{lcl}
 f=f_0+f^{\prime}=
 2\mathcal{J}_x+\Delta-\frac{1}{\beta}\lim\limits_{M\rightarrow \infty}\ln\frac{1}{M}\mathcal{Z}=\\
 2\mathcal{J}_x+\Delta-\frac{1}{\beta}\ln\Lambda_{max},
 \end{array}
 \end{equation}
 \section{Correlation function and thermmal concurrence}\label{Concurrence}
The reduced density matrix of the nearest neighbor sites of the interstitial Heisenberg dimers in the presence of the magnetic field can be expressed as 
\begin{equation}\label{density matrices}
\mathbf{\rho_i} = \left(
\begin{array}{cccc}
u_+ & 0 & 0 & 0 \\
0 & w_1 & z^* & 0 \\
0 & z & w_2 & 0\\
 0 & 0 & 0 & u_-
\end{array} \right),
\end{equation}
where 
\begin{equation}\label{Correlation}
\begin{array}{lcl}
u_{\pm}= \frac{1}{4}(1+\langle\sigma_{i,2}^z\sigma_{i,5}^z\rangle)\pm \frac{\langle\sigma_{i,2}^z+\sigma_{i,5}^z\rangle}{2},\\
w_1=w_2= \frac{1}{4}(1-\langle\sigma_{i,2}^z\sigma_{i,5}^z\rangle),\\
z=z^*=\frac{1}{2}\langle\sigma_{i,2}^x\sigma_{i,5}^x\rangle.
\end{array}
\end{equation}
In the limit of $N\rightarrow \infty $, the correlation functions between Heisenberg spin dimers can be obtained by using a derivative of the free energy with respect to the associated parameters as 
 \begin{equation}\label{Correlationfunctions}
\begin{array}{lcl}
\big\langle\sigma^x_{2}\sigma^x_{5}\big\rangle=-\frac{1}{2}\frac{\partial f}{\partial\mathcal{J}_x},\\
\big\langle\sigma^z_{2}\sigma^z_{5}\big\rangle=-\frac{\partial f}{\partial\Delta},\\
\mathcal{M}_i=\big\langle\sigma^z_i\big\rangle=-\frac{\partial f}{\partial B}.
\end{array}
\end{equation}
Where $\mathcal{M}= \langle\sigma_{2}^z+\sigma_{5}^z\rangle/2$ is the magnetization per block.  
After straightforward algebraic manipulation the concurrence will be as the form \cite{Osterloh1}
\begin{equation}\label{Tm}
\begin{array}{lcl}
\mathcal{C}_{2,5}(i)=
\max\big\lbrace 0,\vert\langle\sigma_{i,2}^x\sigma_{i,5}^x\rangle\vert-
\sqrt{\big(1/2 + 2\langle\sigma_{i,2}^z\sigma_{i,5}^z\rangle\big)^2-\mathcal{M}^2}\big\rbrace
\end{array}
\end{equation}

\begin{figure}
\begin{center}
\includegraphics[width=7.5cm,height=5cm]{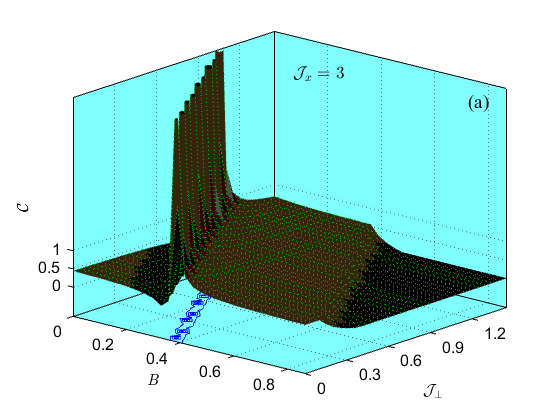}
\includegraphics[width=7.5cm,height=5cm]{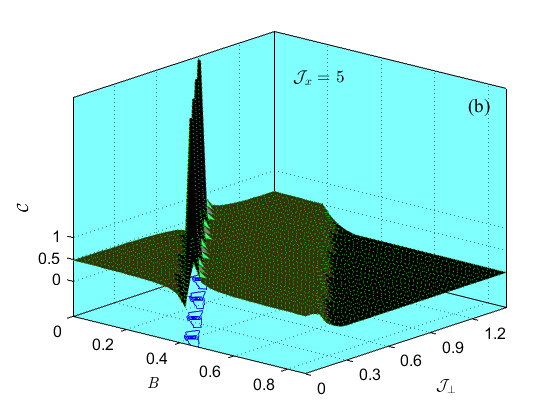}
\includegraphics[width=7.5cm,height=5cm]{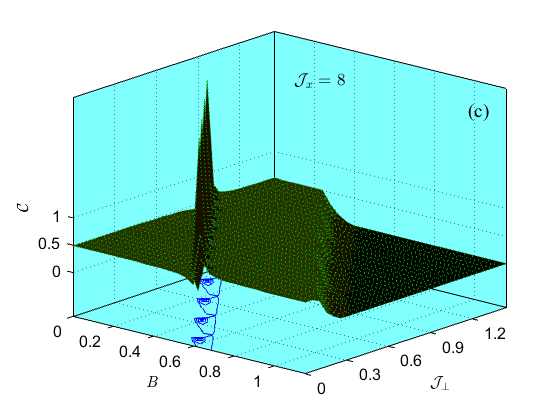}
\caption{Concurrence at low temperature ($\beta=1$) as function of $B$ and $\mathcal{J}_{\perp}$, for fixed values of
 $\mathcal{J}_{\parallel}=K=\mathcal{J}_{Is}=1$ and (a) $\mathcal{J}_{x}=3$, (b) $\mathcal{J}_{x}=5$, and (c) $\mathcal{J}_{x}=8$.}
\label{fig:C1}
\end{center}
\end{figure}
To clarify the effect of  bilinear exchange couplings on the rungs $\mathcal{J}_{\perp}$ upon the  magnetic field  dependence of the bipartite entanglement, we have shown in Fig. \ref{fig:C1} the concurrence as function of the magnetic field and $\mathcal{J}_{\perp}$ at low temperature for the various fixed values of the isotropic parameter $\mathcal{J}_{x}$. As shown in Fig. \ref{fig:C1} (a), at weak magnetic field and  bilinear exchange couplings on the rungs $\mathcal{J}_{\perp}$, the concurrence is not zero. With increase of the magnetic field and 
$\mathcal{J}_{\perp}$, the concurrence decreases because there is a $K$-dependent singularity at which all information about pairwise entanglement obtained by the concurrence destroy. As the magnetic field and $\mathcal{J}_{\perp}$ further increase, the concurrence reaches a smooth plateau for which  $\mathcal{C} < 1$, where the concurrence obtains its normal behavior and we can properly investigate the concurrence and gain knowledge about the pairwise entanglement. Finally, the concurrence decreases and gradually vanishes at stronger magnetic field and $\mathcal{J}_{\perp}$. 

\begin{figure}
\begin{center}
\includegraphics[width=8cm,height=7cm]{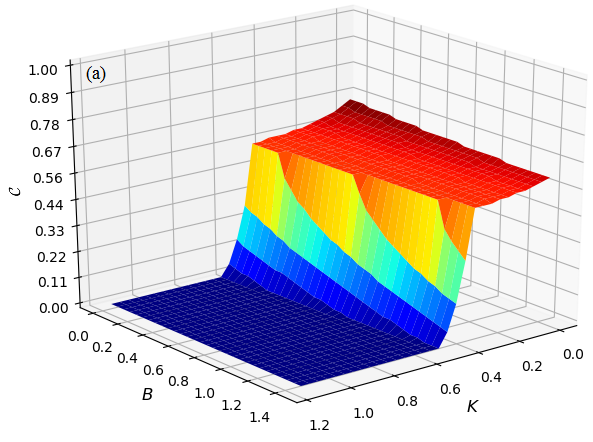}
\includegraphics[width=9cm,height=6cm]{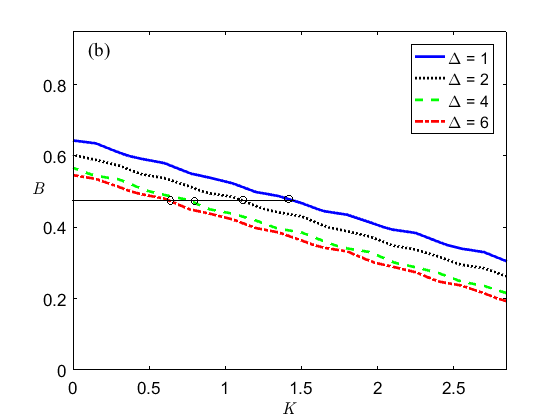}
\caption{Concurrence at low temperature ($\beta=1$) as function of $B$ and $K$, for fixed values of
 $\mathcal{J}_{\parallel}=\mathcal{J}_{\perp}=\Delta=\mathcal{J}_{Is}=1$ and (a) $\mathcal{J}_{x}=3$. (b) The critical points in the ($B-K$) plan at which  the concurrence vanishes for various fixed values of $\Delta$.}
\label{fig:C2}
\end{center}
\end{figure}

By inspecting the blue contour plots in ($B-\mathcal{J}_{\perp}$) plan below the concurrence curves shown in Figs. \ref{fig:C1} (a), (b) and (c), one can see that the region in which singularity occurs moves to stronger magnetic field and  bilinear exchange couplings on the rungs 
$\mathcal{J}_{\perp}$ upon the increasing the isotropic parameter $\mathcal{J}_{x}$. 
Figure \ref{fig:C2} (a) shows the concurrence as function of the magnetic field and  the cyclic four-spin exchange interaction $K$ at low temperature for the fixed values of $\mathcal{J}_{\parallel}=\mathcal{J}_{\perp}=\Delta=\mathcal{J}_{Is}=1$ and $\mathcal{J}_{x}=3$.
It can be seen that, the concurrence exhibits a steep decrease when $K$ gradually increases till it suddenly vanishes at a critical point. As the magnetic field increases, the entanglement vanishing occurs at weaker values of $K$.
To better understand the role of the $K$ in determining appropriate regions in which the concurrence has reasonable behavior, we plot
Fig. \ref{fig:C2} (b) which depicts the critical points in the ($B-K$) plan at which the concurrence vanishes, in this regard various fixed values of the 
$\Delta$ is considered. Interestingly, we see that under the considered conditions, the critical points in the ($B-K$) plan shows different lines upon increasing of the $\Delta$. Indeed by increasing $\Delta$, for a fixed value of the magnetic field, the critical point occurs at weaker cyclic four-spin exchange interaction 
$K$ (horizontal solid line).

In Ref. \cite{Arian1} by investigating the heat capacity function of the fermionic double sawtooth ladder, we noted that the cyclic four-spin exchange interaction $K$ has an essential role to determine the range $\mathcal{J}_{\perp}$ to study the concurrence. Indeed, the singularity was explicitly seen in the heat capacity function and $K$  did not contribute directly in quantum entanglement between interstitial Heisenberg dimer, while here the singularity is interestingly seen in the thermal pairwise concurrence such that  $K$ contributes directly in the thermal entanglement.
Moreover, by comparing Fig. \ref{fig:C2} (b) and the contour lines illustrated in Fig. \ref{fig:C1}, one can precisely detect the singular points.
\section{Conclusions}\label{conclusions}
In this paper, we have examined the thermal concurrence of the mixed spin-(1,1/2) Ising-XXZ double sawtooth ladder consist of the interstitial Heisenberg dimer half-spins connected to the leg sites by using an approximate solution of the transfer-matrix. According to the numerical investigations we understood that the most components of the transfer matrix are almost effectless to derive the largest eigenvalue, so we neglected them and reconstructed a new transfer matrix with less components.  Actually, we used an approximate procedure to solve the model, where it provides the possibility of obtaining numesical and analytical expressions for the thermal concurrence as well as thermodynamic parameters with high accuracy. 

We found that, the concurrence has a strange behavior versus the magnetic field and the bilinear exchange couplings on the rungs 
$\mathcal{J}_{\perp}$. As a matter of fact,  we have seen a special region in which a singularity occurs for the concurrence at low temperature. Interestingly, this singularity is extremely dependent on the cyclic four-spin exchange interaction $K$. Furthermore, we have understood that the concurrence shows a sharp steep decrease upon increasing $K$, where it suddenly vanishes at a critical point. The critical points are dependent on the anisotropic parameter $\Delta$, namely for fixed values of the magnetic field, the critical points occur at  weaker the cyclic four-spin exchange interaction $K$.

We note that the approximate method used in the present paper is valid at low temperature and can be straightforwardly adapted to account for the another spin models with more complicate transfer matrix and different spin arrangement to investigate the thermal pairwise entanglement with high accuracy.

\section*{Acknowledgment}
NA acknowledge by partly financial support of the MC-IRSES no. 612707 (DIONICOS) under FP7-PEOPLE-2013 and ICTP NT-04 grants.

\bibliography{}

\section*{References}

\begin{thebibliography}
\bibliography{}

\bibitem{Horodecki2009}
R.  Horodecki, P.  Horodecki, M.  Horodecki and K.  Horodecki, \href{https://dx.doi.org/10.1103/RevModPhys.81.865}{{ Rev. Mod. Phys.} {\bf 81} (2009) 865-942.}

\bibitem{Wootters1998}
W. K. Wootters, Quantum Inform. Comput.  {{\bf 1} (2001) 27}; W. K. Wootters, Phys. Rev. Lett. \href{https://doi.org/10.1103/PhysRevLett.80.2245}{{\bf 80} (1998) 2245}; 

\bibitem{Zhang2007}
 G. F. Zhang, Physical Review A,  \href{https://doi.org/10.1103/PhysRevA.75.034302}{{\bf 75} (2007) 034304}.
 
 \bibitem{Langari2008}
M. Kargarian, R. Jafari and A. Langari, Phys. rev. A  \href{https://doi.org/10.1103/PhysRevA.77.032346}{{\bf 77} (2008) 032346}.

\bibitem{Saif2010}
A. Saif, M. Hassan, B. Lari and P. S. Joag, J. Phys. A: Math. Theor.  \href{https://doi.org/10.1088/1751-8113/43/48/485302}{{\bf 43} (2010) 485302}.


 \bibitem{Vedral2001}
M. C. Arnesen, S. Bose, V. Vedral, Phys. Rev. Lett. \href{https://dx.doi.org/10.1103/PhysRevLett.87.017901}{ {\bf 87} (2001) 017901}.

 \bibitem{Kamta}
G. L. Kamta and A. F. Starace, Phys. Rev. Lett. \href{https://doi.org/10.1103/PhysRevLett.88.107901}{{\bf 88} (2002) 107901}.

 \bibitem{Ananikian2012}
N. S. Ananikian, L. N. Ananikyan, L. A. Chakhmakhchyan and O. Rojas, {J. Phys.: Condens. Matt.} \href{https://dx.doi.org/10.1088/0953-8984/24/25/256001}{\textbf{24} (2012) 256001}.

\bibitem{Arian2} 
  H. Arian Zad and  H. Movahhedian, \href{https://dx.doi.org/10.1142/S0217979217500941}{{\it Int. J. Mod. Phys. B}  \textbf{31} (2017) 1750094.}

 \bibitem{Werlang2010}
 T. Werlang, C. Trippe, G. A. P. Ribeiro and G. Rigolin, Phys. Rev. Lett. \href{https://dx.doi.org/10.1103/PhysRevLett.105.095702}{{\bf 105} (2010) 095702}.

\bibitem{Abliz2011326}
G. F. Zhang, Z. T. Jiang and A. Abliz, Ann. of Phys.  \href{http://dx.doi.org/10.1016/j.aop.2010.12.005}{{\bf 326} (2011) 867-875}.
  
 \bibitem{Sarandy2012}
 T. R. d. Oliveira, A. Saguia and M. S. Sarandy, Eur. Phys. Lett.  \href{http://dx.doi.org/10.1209/0295-5075/100/60004}{{\bf 100} (2012) 60004}. 
 
 \bibitem{Osterloh1}
 L. Amico, A. Osterloh, F. Plastina, R. Fazio and G. M. Palma,  { Phys. Rev. A} \href{https://dx.doi.org/10.1103/PhysRevA.69.022304}{ \textbf{69} (2004) 022304.}

\bibitem{Furman2012}
 G. B. Furman, V. M. Meerovich and V. L. Sokolovsky, { Phys. Rev. A} \href{https://dx.doi.org/10.1103/PhysRevA.86.032336}{{\bf 86}  (2012)
 032336.}
 
 \bibitem{Dolde}
F. Dolde and I. Jakobi, B. Naydenov {\it et al.}, Nat. comm.  \href{https://dx.doi.org/10.1038/nphys2545}{{\bf 9} (2013) 139.}
 
 \bibitem{ArianNMR}
H. Arian Zad, Commun. Theor. Phys.  {\bf 66} (2016) 629.
 
 \bibitem{Rojas2}
O. Rojas, M. Rojas, N. S. Ananikian and S. M. D. Souza, { Phys. Rev.  A} \href{https://dx.doi.org/10.1103/PhysRevA.86.042330}{ \textbf{86} 
(2012) 042330.}

 \bibitem{Gu}
 B. Gu  and G. Su,  {Phys. Rev. B}  \href{https://dx.doi.org/10.1103/PhysRevB.75.174437}{\textbf{75} ( 2007) 174437}

\bibitem{Abgaryan1}
 V. S. Abgaryan, N. S. Ananikian, L. N. Ananikyan and V. Hovhannisyan, {Solid State Comm.} \href{https://dx.doi.org/10.1016/j.ssc.2014.11.013}{ \textbf{203} (2015) 5.}
 
\bibitem{Abgaryan2}
V. S. Abgaryan, N. S. Ananikian, L. N. Ananikyan and V. Hovhannisyan, {Solid State Comm.} \href{https://dx.doi.org/10.1016/j.ssc.2015.10.003}{ \textbf{224} (2015) 15.}
 
\bibitem{Strecka1}
  O. Rojas,  M. Rojas,  S. M. D. Souza,  J. Torrico,  J. Strecka and  M. L. Lyra, \href{https://dx.doi.org/10.1016/j.jmmm.2016.02.095}{{ Physica  A} {\bf 486} (2017) 367.}
 
\bibitem{Ivanov1} 
N. B. Ivanov  { Condens. Matt. Phys.} \href{https://dx.doi.org/10.5488/CMP.12.3.435}{\textbf{12}  (2009) 435.} 
 
 \bibitem{Strecka}
 J. Strecka,R. C. Alecio, M. Lyra and O. Rojas  {J. Magn. Magn. Mater.} \href{https://dx.doi.org/10.1016/j.jmmm.2016.02.095} {\textbf{409} (2016) 124.}

\bibitem{Buttner}
 S. Chen,  H. Buttner and  J. Voit,   \href{https://doi.org/10.1103/PhysRevB.67.054412}{{ Phys. Rev. B} \textbf{67} (2003) 054412.}

\bibitem{Blundell}
 S. A. Blundell and  M. D. N. Regueir, \href{https://doi.org/10.1140/epjb/e2003-00054-2}{ { Eur. Phys. J. B} \textbf{31} (2003) 453.}

\bibitem{Bacq}
 O. L. Bacq,   A. Pasturel,  C. Lacroix and  M. D. N. Regueiro, \href{https://doi.org/10.1103/PhysRevB.71.014432}{{ Phys. Rev. B} \textbf{71} (2005) 014432.}
 
 \bibitem{Arian1}
 H. Arian Zad and  N. Ananikian,  \href{https://doi.org/10.1088/1361-648X/aa8dd0}{{ J. Phys. Condens. Matt.} \textbf{29}, (2017) 455402.}

 \bibitem{Lisnyi}
B. Lisnyi, J. Strecka, Physica  \href{https://doi.org/10.1016/j.physa.2016.06.088}{ A 462 (2016) 104.}

\bibitem{Verkholyak} 
 J. Strecka and  T. Verkholyak,  \href{https://dx.doi.org/10.1007/s10909-016-1687-5}{ {\it J. Low Temp. Phys.,} {\bf 187} (2017) 712.}

 \bibitem{ArianC2}
 H. Arian Zad and  N. Ananikian, { Accepted for publishing in J. Phys. Condens. Matt.}  (2018).

\bibitem{Paulinelli}
  H. G. Paulinelli,  S. M. de Souza and  O. Rojas,  \href{https://dx.doi.org/10.1088/0953-8984/25/30/306003}{{\it J. Phys. Condens. Matt.}  \textbf{25} (2013) 306003.}


\end{thebibliography}
\end{document}